\documentclass[12pt]{report}
\usepackage[utf8]{inputenc}
\usepackage{blindtext}
\usepackage[letterpaper, total={8.5in, 11in}, margin=2in]{geometry}
\usepackage{ragged2e}
\usepackage{blindtext}
\usepackage{graphicx}
\usepackage{amsmath}
\usepackage{amssymb}
\usepackage{gensymb}
\usepackage{array}
\date{}

\begin{document}
\centering{\huge Analysis of IRS-Assisted NOMA for 6G Wireless Communications\\
\vspace{24pt}
\large Mobasshir Mahbub, Raed M. Shubair}

\newpage

\RaggedRight{\textbf{\Large 1.\hspace{10pt} Introduction}}\\
\vspace{18pt}
\justifying{\noindent For fifth-generation (5G) and beyond wireless communications include indoor localization [1-20], terahertz strategies [21-33], and antenna architecture [34-50] and more expectations on energy usage, spectrum efficiency, and huge connectivity have been put [51]-[53]. IRS technology [54], [55] has recently received a lot of interest. IRS is a metasurface typically made up of numerous passive reflecting elements that allow dynamic modification of signal characteristics. Its capacity to manage signal reflection and change the propagation circumstances can result in a performance boost at reduced power consumption.}

The IRS is envisaged as a game-changing technology for 6G wireless networks. IRS, as opposed to standard wireless relaying technology [56], merely reflects signals and operates in a full-duplex configuration with reduced energy usage. The reflected signal transmission may be jointly adjusted by modifying the phases of the reflecting elements of IRS. This improves throughput, coverage, and energy efficiency.

NOMA [57] has recently gained a lot of interest because of its enormous potential to facilitate vast connections and improve spectrum efficiency. Fig. 1 shows a typical IRS-assisted NOMA communication scenario.

\begin{figure}[htbp]
\centerline{\includegraphics[height=6.5cm, width=10.0cm]{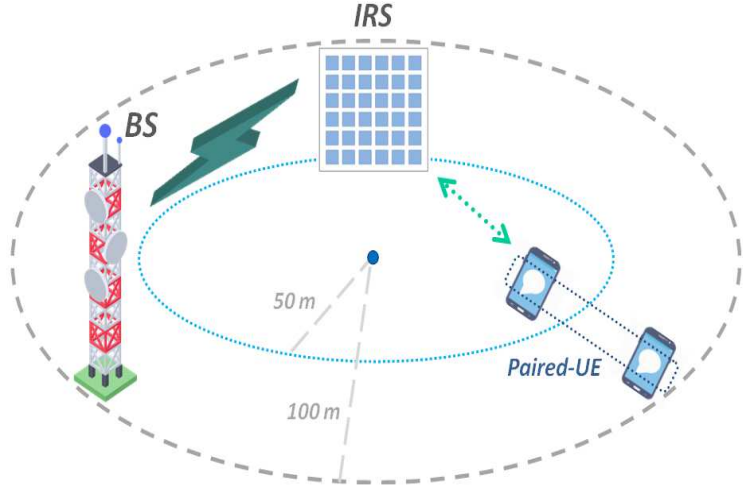}}
\caption{IRS-assisted NOMA communication scenario.}
\label{fig}
\end{figure}

NOMA, as opposed to traditional orthogonal multiple access (OMA) [58], is indeed one of the important innovations in prospective wireless communication networks because of its greater connectivity support, spectral efficiency accomplishment, user fairness guarantee, etc. The fundamental concept of NOMA is to accommodate numerous users over the same resource block (i.e., frequency, time, and code), using superposition coding and successive interference cancellation at the transmitter and receiver end, respectively [59]. Users with superior channel quality are capable of reducing intra-channel interference from users having poor channel conditions.

This article investigated the possible performance enhancement brought about by efficiently combining IRS with NOMA techniques inspired by the aforementioned features of the IRS and NOMA [60].

The work explored an IRS-assisted NOMA downlink system with two users across fading channels i.e. evaluates the performance of the reference model (based on the conventional distance-dependant path loss model) and modified the model by incorporating an IRS-specific frequency-distance-dependent path loss model. Furthermore, this work figure out an improvement scope in the reference model and enhanced the model however the enhancement of the reference model is still unparallel to the modified model.
\vspace{18pt}

\RaggedRight{\textbf{\Large 2.\hspace{10pt} Related Literature}}\\
\vspace{18pt}
\justifying{\noindent The section of the paper briefed several prior works and literature relative to the downlink IRS-assisted NOMA communication.

Wang et al. [61] examined the effectiveness of the deployment of IRS-assisted NOMA in terms of transmit power consumption. In this context, the downlink transmission power minimization problem is analyzed considering the constraint of the user's minimum level of SINR. Cheng et al. [62] studied a downlink NOMA system, in which an IRS is deployed to improve the network coverage by supporting cell-edge devices. Fu et al. [63] considered the downlink transmit power optimization problem for an IRS-assisted NOMA system by optimizing the transmit beamforming and the phase-shift matrix. Zhu et al. [64] proposed and analyzed a downlink multi-input single-output (MISO) NOMA scheme assisted by an IRS. Sena et al. [65] exploited dual-polarized IRS to improve the performance of massive multi-input multi-output (MIMO) NOMA in the case of imperfect successive interference cancellation (SIC). Zuo et al. [66] investigated the downlink IRS-aided NOMA to maximize the network throughput in terms of resource allocation namely the channel assignment, power allocation, and reflection coefficients. Liu et al. [67] proposed and analyzed a millimeter-wave (mmWave) downlink IRS-assisted massive MIMO NOMA system incorporating a lens antenna array under transmission power constraints.
}
\vspace{18pt}

\newpage
\RaggedRight{\textbf{\Large 3.\hspace{10pt} System Model}}\\
\vspace{18pt}
\justifying{\noindent The research considered an IRS-empowered downlink NOMA system in which the base station and paired user equipment both are performing communication with a single antenna. In which a near and a far user (an outer circle user having separation distance twice of the near user from the IRS) denoted by $\mathsf{u}_1$ and $\mathsf{u}_2$ respectively are assigned to be operated in a pair to share the common frequency-time resource block.

The base station transmits the superimposed signal of the users $x = \sqrt{\wp}(a_1 \mathsf{s}_1+ a_2 \mathsf{s}_2)$, where $\wp$ is the transmission power of the base station, $\mathsf{s}_1$ and $\mathsf{s}_2$ are the complex message of the near and far user $\mathsf{u}_1$ and $\mathsf{u}_2$ respectively. During the transmission, the SIC technique is utilized at the receiver end, therefore superposition coding (SC) is applied at the base station, and a higher power coefficient is assigned to a far user due to the weaker channel characteristics (i.e. highly faded channel), namely $a_2  > a_1$ and $a_1^2+a_2^2=1$. According to the principle of NOMA, the far user $\mathsf{u}_2$ have to decode its message or signal $\mathsf{s}_2$ considering the message of the near user $\mathsf{u}_1$ as interference.}

\vspace{12pt}
\RaggedRight{\textit{\large A.\hspace{10pt} Conventional Model}}\\
\vspace{12pt}

\justifying The received signal at the user-end $(\mathsf{u}_i)$ is given by (Eq. 1) [68],

\begin{equation}
y_i= \frac{g_i^{T}\mathbf{\Theta} g_0}{\sqrt{\mathcal{L}(d)_{BS-IRS}\mathcal{L}(d)_{IRS-\mathsf{u}_i}}} x + \mathsf{w}
\end{equation}
where $\mathbf{\Theta} = diag(e^{j\theta_1 },e^{j\theta_2 },...,e^{\theta_N}) \in \mathbb{C}^{N\times N}$ is the diagonal phase shift matrix of the IRS. $g_i \in \mathbb{C}^{M\times 1}$ defines the channel fading between the user $(\mathsf{u}_i)$ and the IRS. $g_0 \in \mathbb{C}^{M\times 1}$ indicates the fading between the base station and the IRS. $\mathcal{L}(d)[dB]= 35.1+36.7log_{10} (d)-\mathsf{G}_t-\mathsf{G}_r$ [18] is the formula for measuring the path loss of the communication channels, $\mathsf{G}_t=10$ dB and $\mathsf{G}_r=10$ dB [68] are the transmitter and receiver gains, respectively, and $d$ denotes the separation distance for either base and IRS or IRS and user. $\mathsf{w}$ is the Gaussian noise (-94 dBm).

Therefore, the received SINR for the user $\mathsf{u}_2$ can be formulated by (Eq. 2),

\begin{equation}
\mathfrak{S}_2= \frac{\frac{|g_2^{T}\mathbf{\Theta} g_0|^2\wp a_2^2}{\mathcal{L}(d)_{BS-IRS}\mathcal{L}(d)_{IRS-\mathsf{u}_2}}}{\frac{|g_1^{T}\mathbf{\Theta} g_0|^2\wp a_1^2}{\mathcal{L}(d)_{BS-IRS}\mathcal{L}(d)_{IRS-\mathsf{u}_1}}+\mathsf{w}}
\end{equation}

After decoding the received SINR of $\mathsf{u}_2$ the received SNR for the user $\mathsf{u}_1$ can be measured by (Eq. 3),

\begin{equation}
\mathfrak{S}_1= \frac{\frac{|g_1^{T}\mathbf{\Theta} g_0|^2\wp a_1^2}{\mathcal{L}(d)_{BS-IRS}\mathcal{L}(d)_{IRS-\mathsf{u}_1}}}{\mathsf{w}}
\end{equation}

\newpage
\vspace{12pt}
\RaggedRight{\textit{\large B.\hspace{10pt} Modified Model}}\\
\vspace{12pt}

\justifying The research modified the [69] reference model by incorporating a dedicated frequency-distance-dependent path loss model for IRS-assisted communication instead of a typical or conventional distance-dependant path loss model.

The received signal at the user-end is formulated by (Eq. 4) incorporating (Eq. 5),

\begin{equation}
y_i= \frac{g_i^{T}\mathbf{\Theta} g_0}{\sqrt{\mathcal{L}_{IRS(\mathsf{u}_i)}}} x + \mathsf{w}
\end{equation}
where
\begin{equation}
\mathcal{L}_{IRS} = \frac{64\pi^3(d_1 d_2)^2}{M^2 N^2 \lambda^2 A^2 \mathsf{G} \mathsf{G}_t \mathsf{G}_r d_x d_y cos(\theta_t) cos(\theta_r)}
\end{equation}
where
\begin{equation*}
d_1 = \sqrt{(x^{BS}-x^{IRS})^2+(y^{BS}-y^{IRS})^2+(z^{BS}-z^{IRS})^2}
\end{equation*}
is the separation between the base station and  the IRS positioned at $(x^{BS},y^{BS},z^{BS})$ and $(x^{IRS},y^{IRS},z^{IRS})$ respectively. 
\begin{equation*}
d_2 = \sqrt{(x^{IRS}-x^{U})^2+(y^{IRS}-y^{U})^2+(z^{IRS}-z^{U})^2}
\end{equation*}
indicates the separation distance between the user located at $(x^U,y^U,z^U)$ and the IRS. The transmitter and receiver gains are $\mathsf{G}_t$ and $\mathsf{G}_r$. The scattering gain is determined by $\mathsf{G} = \frac{4\pi d_x d_y}{\lambda^2}$. The numbers of transmitting and receiving elements are denoted by $M$ and $N$, respectively. $d_x$ is the length and $d_y$ is the width of the elements of IRS. The carrier wavelength is $\lambda$. $\theta_t$ and $\theta_r$ are the transmitting and  receiving angles. The reflection coefficient of the IRS is $A$.

The received SINR at $\mathsf{u}_2$ is given by (Eq. 6),

\begin{equation}
\mathfrak{S}_2= \frac{\frac{|g_2^{T}\mathbf{\Theta} g_0|^2\wp a_2^2}{\mathcal{L}_{IRS(\mathsf{u}_2)}}}{\frac{|g_1^{T}\mathbf{\Theta} g_0|^2\wp a_1^2}{\mathcal{L}_{IRS(\mathsf{u}_1)}}+\mathsf{w}}
\end{equation}

The received SNR at $\mathsf{u}_1$ is measured by (Eq. 7),

\begin{equation}
\mathfrak{S}_1= \frac{\frac{|g_1^{T}\mathbf{\Theta} g_0|^2\wp a_1^2}{\mathcal{L}_{IRS(\mathsf{u}_1)}}}{\mathsf{w}}
\end{equation}

\vspace{18pt}

\RaggedRight{\textbf{\Large 4.\hspace{10pt} Numerical Results and Discussions}}\\
\vspace{18pt}
\justifying{\noindent The section of the paper includes the measurement results and corresponding discussions on the derived results. Table I includes the measurement parameters and values.

\begin{table}[htbp]
\caption{Parameter and Values}
\begin{center}
\begin{tabular}{| m{3.5cm} | m{3.5cm}|}
\hline
\textbf{\textit{Parameters}}& \textbf{\textit{Values}}\\
\hline
Cell area & 200x200m\\
\hline
Transmit power & 6W \\
\hline
Transmitter and receiver gain & 5 dB (This work),
10 dB [18], 20 dB (Enhanced [20] for [18])\\
\hline
Number of IRS transmit-receive elements & 64\\
\hline
Length and width of the IRS elements & 0.0038m\\
\hline
Transmit and receive angle & 45\degree \\
\hline
Reflection coefficient & 0.9\\
\hline
Carrier frequency & 90 GHz\\
\hline
\end{tabular}
\label{tab1}
\end{center}
\end{table}

Fig. 2 shows the measurement of received power for both the models in the context of user 1.

\begin{figure}[htbp]
\centerline{\includegraphics[height=8.5cm, width=11.5cm]{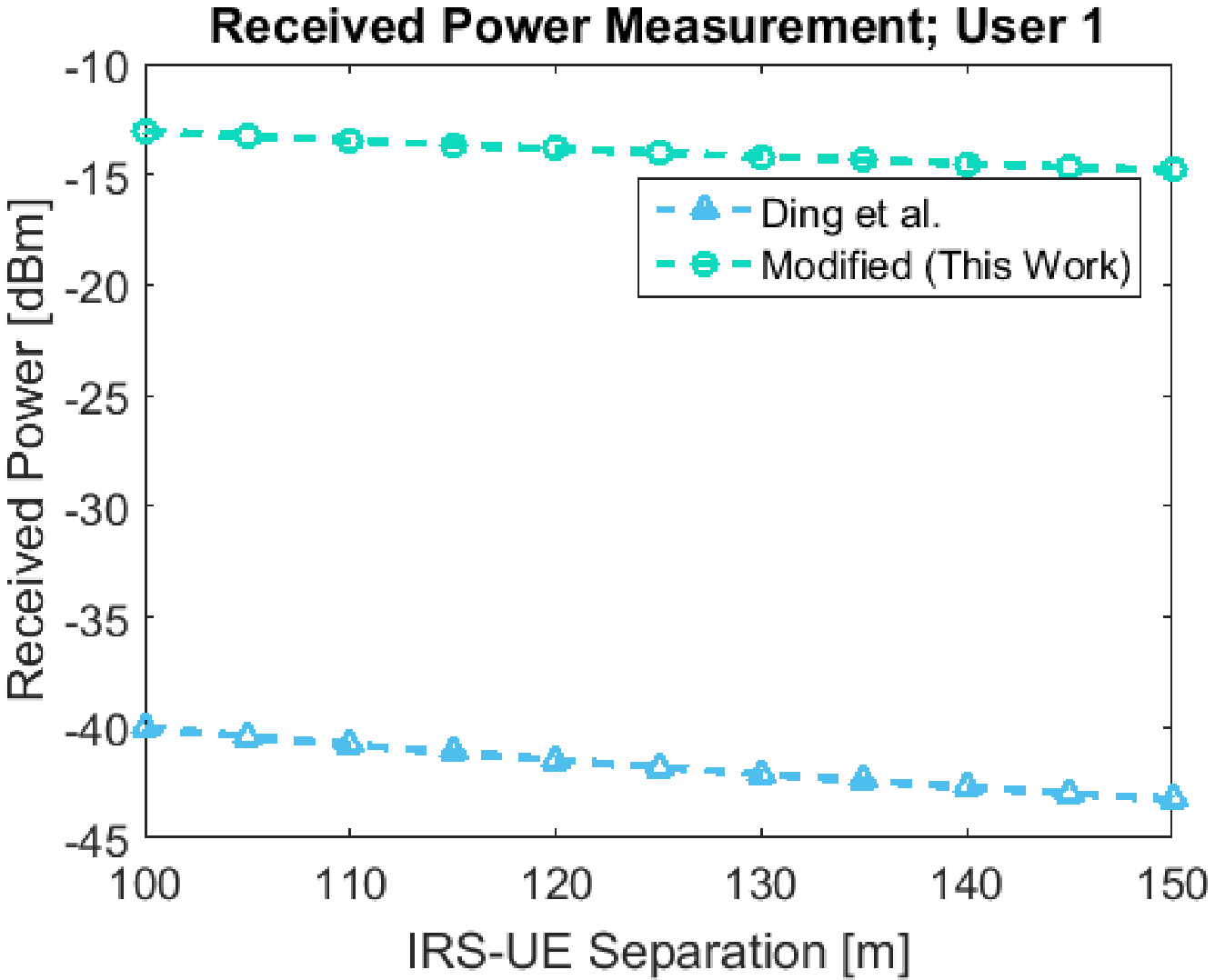}}
\caption{Received power at user 1.}
\label{fig}
\end{figure}

Fig. 3 illustrates the measurement of received power for both of the models in the context of user 2.

\begin{figure}[htbp]
\centerline{\includegraphics[height=8.5cm, width=11.5cm]{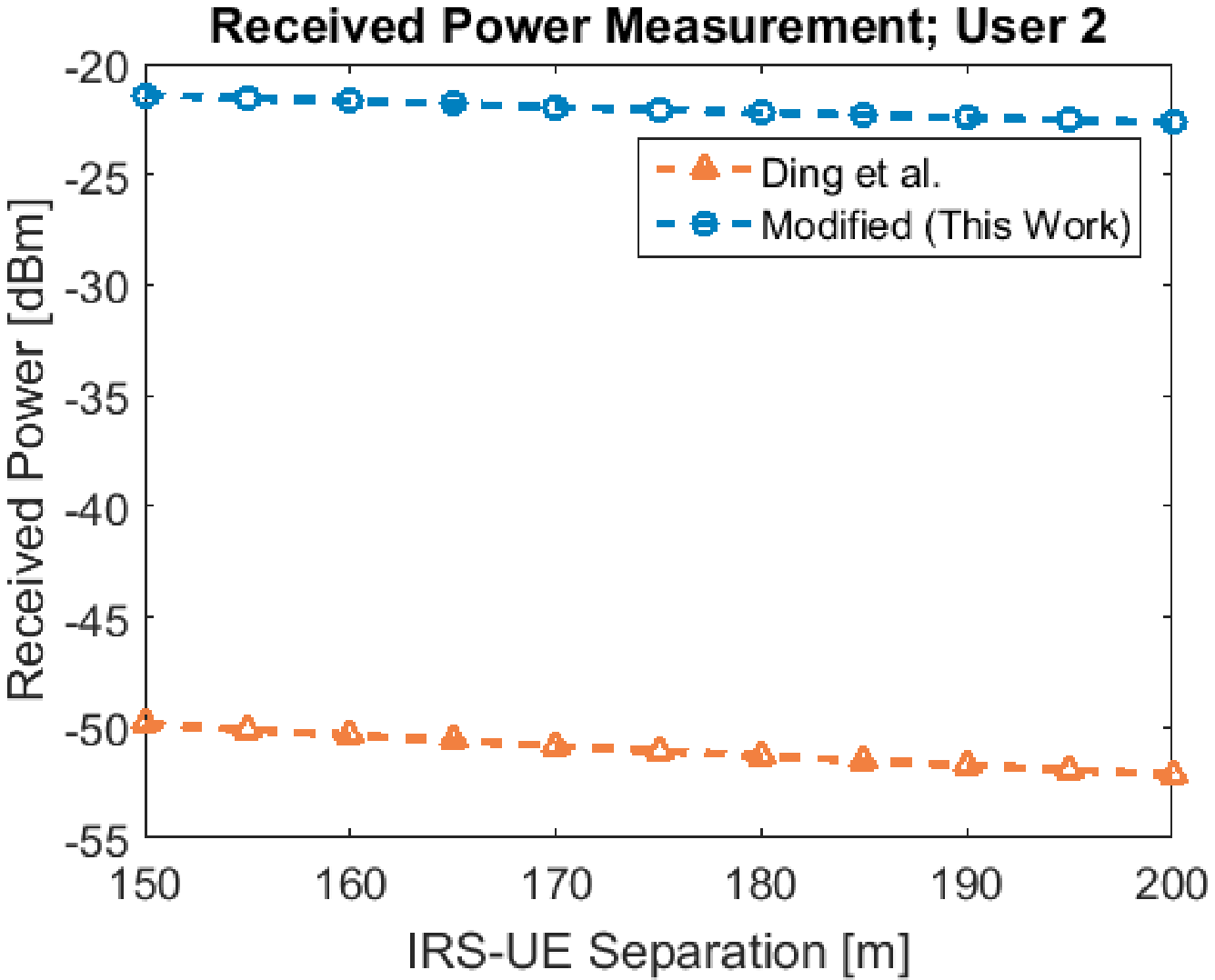}}
\caption{Received power at user 2.}
\label{fig}
\end{figure}

Fig. 4 represents the SINR measurement in the case of user 1 for both of the models.

\begin{figure}[htbp]
\centerline{\includegraphics[height=8.5cm, width=11.5cm]{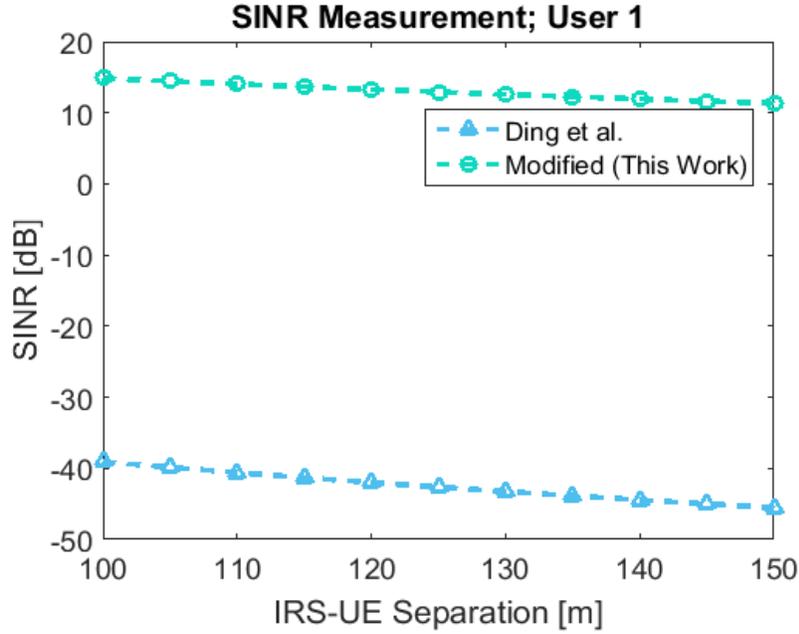}}
\caption{SINR at user 1.}
\label{fig}
\end{figure}

Fig. 5 visualizes the SINR measurement in the case of user 2 for both of the models.

\begin{figure}[htbp]
\centerline{\includegraphics[height=8.5cm, width=11.5cm]{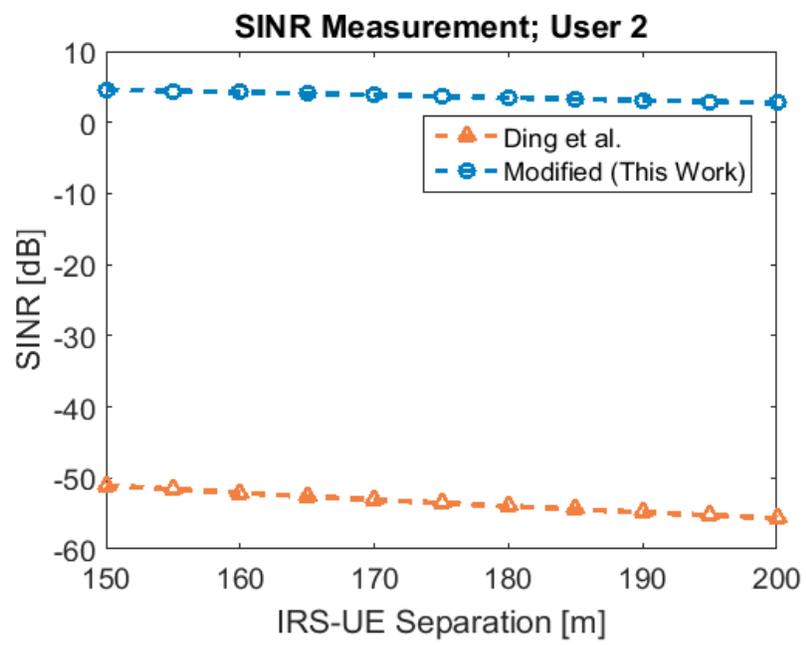}}
\caption{SINR at user 2.}
\label{fig}
\end{figure}

Fig. 6 shows the measurement of received power in the context of user 1 increasing the transmitter-receiver gain considered in the research of Ding et al. [68] through the utilization of an mmWave horn antenna [70].

\begin{figure}[htbp]
\centerline{\includegraphics[height=8.5cm, width=11.5cm]{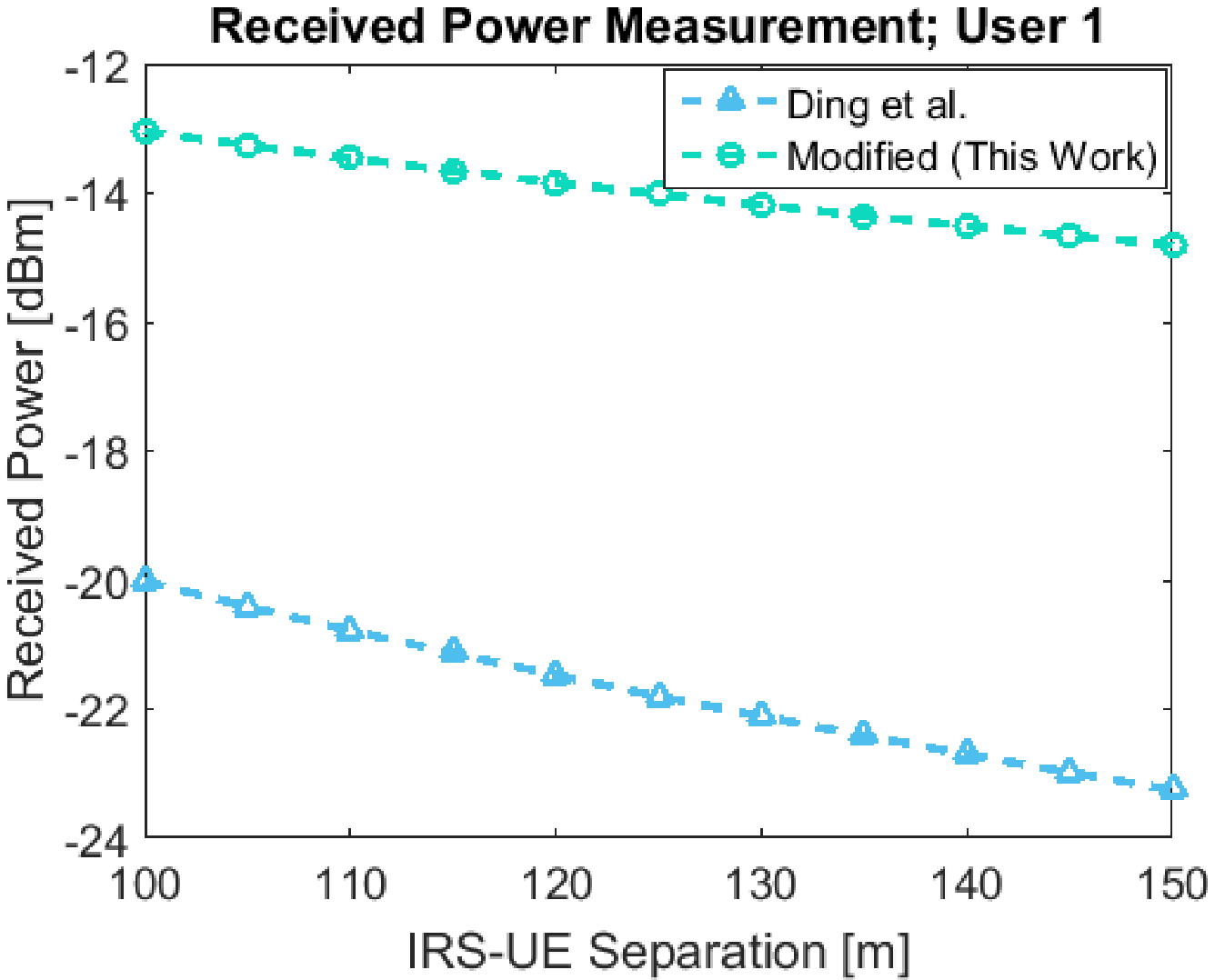}}
\caption{Received power at user 1 (utilizing horn antenna at Ding et al. model).}
\label{fig}
\end{figure}

Fig. 7 illustrates the received power for user 2 increasing the transmitter-receiver gain [70] considered in the research of Ding et al. [68].

\begin{figure}[htbp]
\centerline{\includegraphics[height=8.5cm, width=11.5cm]{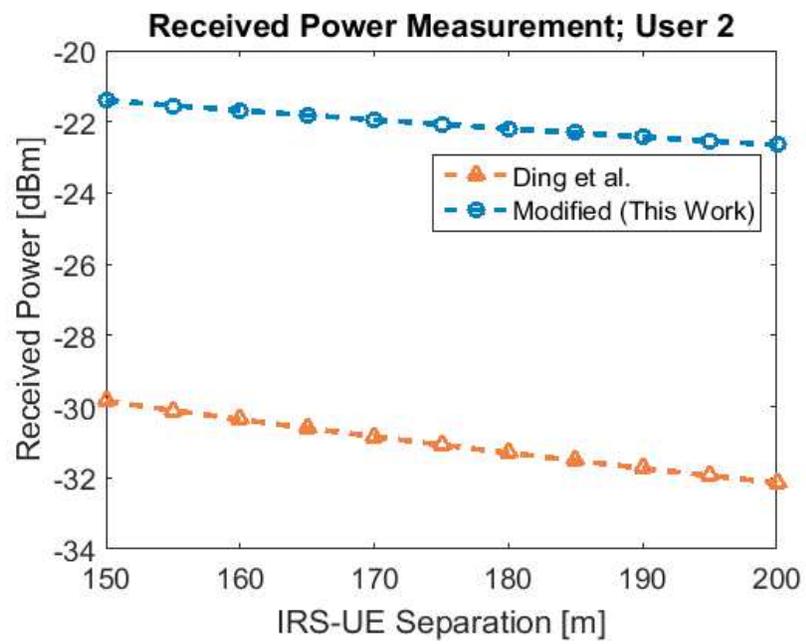}}
\caption{Received power at user 2 (utilizing horn antenna at Ding et al. model).}
\label{fig}
\end{figure}

Fig. 8 represents the measurement of SINR of user 1 for both the models increasing the transmitter-receiver gain through the utilization of an mmWave horn antenna [70] in the Ding et al. [68] model.

\begin{figure}[htbp]
\centerline{\includegraphics[height=8.5cm, width=11.5cm]{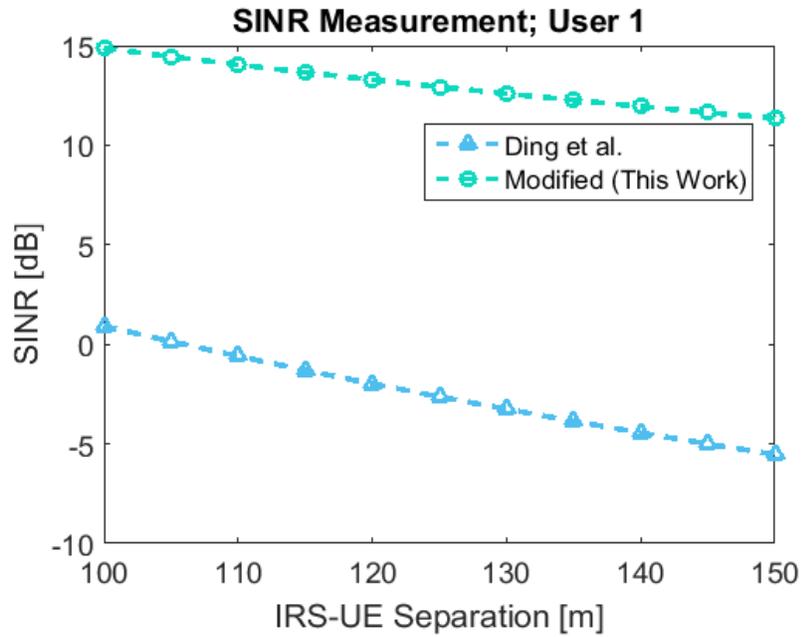}}
\caption{SINR at user 1 (utilizing horn antenna at Ding et al. model).}
\label{fig}
\end{figure}

Fig. 9 shows the measurement of SINR of user 2 increasing the transmitter-receiver gain through the utilization of an mmWave horn antenna in the Ding et al. [68] model.

\begin{figure}[htbp]
\centerline{\includegraphics[height=8.5cm, width=11.5cm]{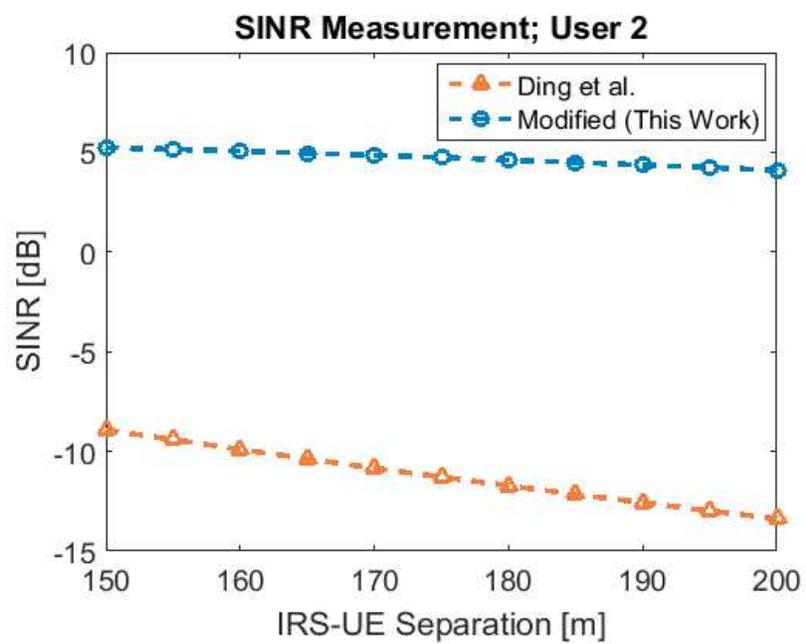}}
\caption{SINR at user 2 (utilizing horn antenna at Ding et al. model).}
\label{fig}
\end{figure}

According to the observation of Figs. 2-5 it is evident that the modified model incorporating a dedicated path loss model for IRS ensures better measurements of received power and SINR compared to the conventional path loss equation-based model presented in the work of Ding et al. [68].

Further, the work tried to enhance the performance of the model presented in the work of Ding et al. [68]. In this circumstance, the work considers the gain of a horn antenna in the conventional path loss model included in the work of Ding et al. [68]. The consideration of the enhanced gain achievable by mmWave horn antenna [70] enhances the performance of the model of Ding et al. [68] but is still unparalleled to the modified model presented by this work even with a lower level of transmitter-receiver gain (as per the observation of Figs. 6-9).

The work modified the [68] reference model incorporating a dedicated frequency-distance-dependent path loss model for IRS-assisted communication instead of a typical or conventional distance-dependant path loss model. From the point of view of this research, since the reference work is on the IRS-assisted NOMA system an IRS-specific path loss model should be preferable to figure out a more convenient measurement result. As per the observation of this work the conventional path loss model is unable to provide the realistic measurement result as an IRS-specific model.}

\vspace{18pt}

\RaggedRight{\textbf{\Large 5.\hspace{10pt} Conclusion}}\\
\vspace{18pt}
\justifying \noindent {The research targeted the investigation of the IRS-enhanced NOMA scheme for downlink. Relative literature and works are reviewed to obtain an insight into recent developments and figure out further enhancement scopes. According to the observation, it seemed that it will be more convenient to adopt an IRS-specific path loss model instead of a conventional path loss model. This work, therefore, modified the reference model by adopting an IRS-specific path loss model instead of a typical or conventional path loss model. The research derived that the incorporation of an IRS-specific model provides a more convenient measurement compared to the conventional model. Moreover, the work figures out an enhancement scope in the reference model by considering the gain of the horn antenna. The approach enhances the performance of the reference model but still, it’s unparalleled to the model modified by this work. The research will be supportive to the researchers and enthusiasts for performing extended research on the relative research issue.}
\vspace{18pt}

\RaggedRight{\textbf{\Large References}}\\
\vspace{12pt}

\justifying{
1.	Mohamed I. AlHajri, Nazar T. Ali, and Raed M. Shubair. "Classification of indoor environments for IoT applications: A machine learning approach." IEEE Antennas and Wireless Propagation Letters 17, no. 12 (2018): 2164-2168.

2.	Mohamed I. AlHajri, Nazar T. Ali, and Raed M. Shubair. "Indoor localization for IoT using adaptive feature selection: A cascaded machine learning approach." IEEE Antennas and Wireless Propagation Letters 18, no. 11 (2019): 2306-2310.

3.	Fahad Belhoul, Raed M. Shubair, and Mohammed E. Al-Mualla. "Modelling and performance analysis of DOA estimation in adaptive signal processing arrays." In ICECS, pp. 340-343. 2003.

4.	Ebrahim M. Al-Ardi, Raed M. Shubair, and Mohammed E. Al-Mualla. "Direction of arrival estimation in a multipath environment: An overview and a new contribution." Applied Computational Electromagnetics Society Journal 21, no. 3 (2006): 226.

5.	R. M. Shubair and A. Al-Merri. "Robust algorithms for direction finding and adaptive beamforming: performance and optimization." In The 2004 47th Midwest Symposium on Circuits and Systems, 2004. MWSCAS'04., vol. 2, pp. II-II. IEEE, 2004.

6.	R. M. Shubair and Y. L. Chow. "A closed-form solution of vertical dipole antennas above a dielectric half-space." IEEE transactions on antennas and propagation 41, no. 12 (1993): 1737-1741.

7.	R. M. Shubair and W. Jessmi. "Performance analysis of SMI adaptive beamforming arrays for smart antenna systems." In 2005 IEEE Antennas and Propagation Society International Symposium, vol. 1, pp. 311-314. IEEE, 2005.

8.	E. M. Al-Ardi, Raed M. Shubair, and M. E. Al-Mualla. "Performance evaluation of direction finding algorithms for adapative antenna arrays." In 10th IEEE International Conference on Electronics, Circuits and Systems, 2003. ICECS 2003. Proceedings of the 2003, vol. 2, pp. 735-738. IEEE, 2003.

9.	Mohamed AlHajri, Abdulrahman Goian, Muna Darweesh, Rashid AlMemari, Raed Shubair, Luis Weruaga, and Ahmed AlTunaiji. "Accurate and robust localization techniques for wireless sensor networks." arXiv preprint arXiv:1806.05765 (2018).

10.	Raed M. Shubair, and Ali Hakam. "Adaptive beamforming using variable step-size LMS algorithm with novel ULA array configuration." In 2013 15th IEEE International Conference on Communication Technology, pp. 650-654. IEEE, 2013.

11.	Zhenghua Chen, Mohamed I. AlHajri, Min Wu, Nazar T. Ali, and Raed M. Shubair. "A novel real-time deep learning approach for indoor localization based on RF environment identification." IEEE Sensors Letters 4, no. 6 (2020): 1-4.

12.	Mohamed I. AlHajri, Nazar T. Ali, and Raed M. Shubair. "A machine learning approach for the classification of indoor environments using RF signatures." In 2018 IEEE Global Conference on Signal and Information Processing (GlobalSIP), pp. 1060-1062. IEEE, 2018.

13.	Satish R. Jondhale, Raed Shubair, Rekha P. Labade, Jaime Lloret, and Pramod R. Gunjal. "Application of supervised learning approach for target localization in wireless sensor network." In Handbook of Wireless Sensor Networks: Issues and Challenges in Current Scenario's, pp. 493-519. Springer, Cham, 2020.

14.	Raed M. Shubair, Abdulrahman S. Goian, Mohamed I. AlHajri, and Ahmed R. Kulaib. "A new technique for UCA-based DOA estimation of coherent signals." In 2016 16th Mediterranean Microwave Symposium (MMS), pp. 1-3. IEEE, 2016.

15.	W. Njima, M. Chafii, A. Chorti, Raed M. Shubair, and H. Vincent Poor. "Indoor localization using data augmentation via selective generative adversarial networks." IEEE Access 9 (2021): 98337-98347.

16.	R. M. Shubair, and A. Merri. "A convergence study of adaptive beamforming algorithms used in smart antenna systems." In 11th International Symposium on Antenna Technology and Applied Electromagnetics [ANTEM 2005], pp. 1-5. IEEE, 2005.

17.	Mohamed Ibrahim Alhajri, N. T. Ali, and R. M. Shubair. "2.4 ghz indoor channel measurements." IEEE Dataport (2018).

18.	M. I. AlHajri, N. T. Ali, and R. M. Shubair. "2.4 ghz indoor channel measurements data set." UCI Machine Learning Repository (2018).

19.	WafaNjima, Marwa Chafii, and Raed M. Shubair. "Gan based data augmentation for indoor localization using labeled and unlabeled data." In 2021 International Balkan Conference on Communications and Networking (BalkanCom), pp. 36-39. IEEE, 2021.

20.	M. I. AlHajri, R. M. Shubair, and M. Chafii. "Indoor Localization Under Limited Measurements: A Cross-Environment Joint Semi-Supervised and Transfer Learning Approach." In 2021 IEEE 22nd International Workshop on Signal Processing Advances in Wireless Communications (SPAWC), pp. 266-270. IEEE, 2021.

21.	Hadeel Elayan, Raed M. Shubair, and Asimina Kiourti. "Wireless sensors for medical applications: Current status and future challenges." In 2017 11th European Conference on Antennas and Propagation (EUCAP), pp. 2478-2482. IEEE, 2017.

22.	Hadeel Elayan, Raed M. Shubair, and Asimina Kiourti. "Wireless sensors for medical applications: Current status and future challenges." In 2017 11th European Conference on Antennas and Propagation (EUCAP), pp. 2478-2482. IEEE, 2017.

23.	Hadeel Elayan, Pedram Johari, Raed M. Shubair, and Josep Miquel Jornet. "Photothermal modeling and analysis of intrabody terahertz nanoscale communication." IEEE transactions on nanobioscience 16, no. 8 (2017): 755-763.

24.	Samar Elmeadawy and Raed M. Shubair. "6G wireless communications: Future technologies and research challenges." In 2019 international conference on electrical and computing technologies and applications (ICECTA), pp. 1-5. IEEE, 2019.

25.	Hadeel Elayan, Raed M. Shubair, and Asimina Kiourti. "On graphene-based THz plasmonic nano-antennas." In 2016 16th mediterranean microwave symposium (MMS), pp. 1-3. IEEE, 2016.

26.	Hadeel Elayan, Raed M. Shubair, Akram Alomainy, and Ke Yang. "In-vivo terahertz em channel characterization for nano-communications in wbans." In 2016 IEEE International Symposium on Antennas and Propagation (APSURSI), pp. 979-980. IEEE, 2016.

27.	Hadeel Elayan, and Raed M. Shubair. "On channel characterization in human body communication for medical monitoring systems." In 2016 17th International Symposium on Antenna Technology and Applied Electromagnetics (ANTEM), pp. 1-2. IEEE, 2016.

28.	Hadeel Elayan, Raed M. Shubair, and Josep M. Jornet. "Characterising THz propagation and intrabody thermal absorption in iWNSNs." IET Microwaves, Antennas \& Propagation 12, no. 4 (2018): 525-532.

29.	Mayar Lotfy, Raed M. Shubair, Nassir Navab, and Shadi Albarqouni. "Investigation of focal loss in deep learning models for femur fractures classification." In 2019 International Conference on Electrical and Computing Technologies and Applications (ICECTA), pp. 1-4. IEEE, 2019.

30.	Hadeel Elayan, Cesare Stefanini, Raed M. Shubair, and Josep M. Jornet. "Stochastic noise model for intra-body terahertz nanoscale communication." In Proceedings of the 5th ACM International Conference on Nanoscale Computing and Communication, pp. 1-6. 2018.

31.	Abdul Karim Gizzini, Marwa Chafii, Ahmad Nimr, Raed M. Shubair, and Gerhard Fettweis. "Cnn aided weighted interpolation for channel estimation in vehicular communications." IEEE Transactions on Vehicular Technology 70, no. 12 (2021): 12796-12811.

32.	Hadeel Elayan, Raed M. Shubair, Josep M. Jornet, Asimina Kiourti, and Raj Mittra. "Graphene-Based Spiral Nanoantenna for Intrabody Communication at Terahertz." In 2018 IEEE Intl. Symposium on Antennas and Propagation \& USNC/URSI National Radio Science Meeting, pp. 799-800. IEEE, 2018.

33.	Abdul Karim Gizzini, Marwa Chafii, Shahab Ehsanfar, and Raed M. Shubair. "Temporal Averaging LSTM-based Channel Estimation Scheme for IEEE 802.11 p Standard." arXiv preprint arXiv:2106.04829 (2021).

34.	Ahmed A. Ibrahim,  JanMachac, and Raed M. Shubair. "Compact UWB MIMO antenna with pattern diversity and band rejection characteristics." Microwave and Optical Technology Letters 59, no. 6 (2017): 1460-1464.

35.	M. Saeeed Khan, Adnan Iftikhar, Sajid M. Asif, Antonio‐Daniele Capobianco, and Benjamin D. Braaten. "A compact four elements UWB MIMO antenna with on‐demand WLAN rejection." Microwave and Optical Technology Letters 58, no. 2 (2016): 270-276.

36.	M. S. Khan, Adnan Iftikhar, Antonio‐Daniele Capobianco, Raed M. Shubair, and Bilal Ijaz. "Pattern and frequency reconfiguration of patch antenna using PIN diodes." Microwave and Optical Technology Letters 59, no. 9 (2017): 2180-2185.

37.	A. Omar, and R. Shubair. "UWB coplanar waveguide-fed-coplanar strips spiral antenna." In 2016 10th European Conference on Antennas and Propagation (EuCAP), pp. 1-2. IEEE, 2016.

38.	Raed M. Shubair, Amna M. AlShamsi, Kinda Khalaf, and Asimina Kiourti. "Novel miniature wearable microstrip antennas for ISM-band biomedical telemetry." In 2015 Loughborough Antennas \& Propagation Conference (LAPC), pp. 1-4. IEEE, 2015.

39.	Muhammad S. Khan, Syed A. Naqvi, Adnan Iftikhar, Sajid M. Asif, Adnan Fida, and Raed M. Shubair. "A WLAN band‐notched compact four element UWB MIMO antenna." International Journal of RF and Microwave Computer‐Aided Engineering 30, no. 9 (2020): e22282.

40.	M. S. Khan, F. Rigobello, Bilal Ijaz, E. Autizi, A. D. Capobianco, R. Shubair, and S. A. Khan. "Compact 3‐D eight elements UWB‐MIMO array." Microwave and Optical Technology Letters 60, no. 8 (2018): 1967-1971.

41.	Mohammed S Al-Basheir, Raed M Shubai, and Sami M. Sharif. "Measurements and analysis for signal attenuation through date palm trees at 2.1 GHz frequency." (2006).

42.	M. S. Khan, A. Iftikhar, Raed M. Shubair, Antonio-Daniele Capobianco, Sajid Mehmood Asif, Benjamin D. Braaten, and Dimitris E. Anagnostou. "Ultra-compact reconfigurable band reject UWB MIMO antenna with four radiators." Electronics 9, no. 4 (2020): 584.

43.	Muhammad S. Khan, Adnan Iftikhar, Raed M. Shubair, Antonio D. Capobianco, Benjamin D. Braaten, and Dimitris E. Anagnostou. "A four element, planar, compact UWB MIMO antenna with WLAN band rejection capabilities." Microwave and Optical Technology Letters 62, no. 10 (2020): 3124-3131.

44.	Hari Shankar Singh, SachinKalraiya, Manoj Kumar Meshram, and Raed M. Shubair. "Metamaterial inspired CPW‐fed compact antenna for ultrawide band applications." International Journal of RF and Microwave Computer‐Aided Engineering 29, no. 8 (2019): e21768.

45.	Raed M. Shubair, Amer Salah, and Alaa K. Abbas. "Novel implantable miniaturized circular microstrip antenna for biomedical telemetry." In 2015 IEEE International Symposium on Antennas and Propagation \& USNC/URSI National Radio Science Meeting, pp. 947-948. IEEE, 2015.

46.	Yazan Al-Alem, Ahmed A. Kishk, and Raed M. Shubair. "Enhanced wireless interchip communication performance using symmetrical layers and soft/hard surface concepts." IEEE Transactions on Microwave Theory and Techniques 68, no. 1 (2019): 39-50.

47.	Yazan Al-Alem, Raed M. Shubair, and Ahmed Kishk. "Efficient on-chip antenna design based on symmetrical layers for multipath interference cancellation." In 2016 16th Mediterranean Microwave Symposium (MMS), pp. 1-3. IEEE, 2016.

48.	Asimina Kiourti, and Raed M. Shubair. "Implantable and ingestible sensors for wireless physiological monitoring: a review." In 2017 IEEE International Symposium on Antennas and Propagation \& USNC/URSI National Radio Science Meeting, pp. 1677-1678. IEEE, 2017.

49.	Melissa Eugenia Diago-Mosquera, Alejandro Aragón-Zavala, Fidel Alejandro Rodríguez-Corbo, Mikel Celaya-Echarri, Raed M. Shubair, and Leyre Azpilicueta. "Tuning Selection Impact on Kriging-Aided In-Building Path Loss Modeling." IEEE Antennas and Wireless Propagation Letters 21, no. 1 (2021): 84-88.

50.	Yazan Al-Alem, Ahmed A. Kishk, and Raed Shubair. "Wireless chip to chip communication link budget enhancement using hard/soft surfaces." In 2018 IEEE Global Conference on Signal and Information Processing (GlobalSIP), pp. 1013-1014. IEEE, 2018.

51. A. Dogra, R. K. Jha and S. Jain, "A Survey on Beyond 5G Network With the Advent of 6G: Architecture and Emerging Technologies," in \textit{IEEE Access}, vol. 9, pp. 67512-67547, 2021.

52. S. Elmeadawy and R. M. Shubair, "6G Wireless Communications: Future Technologies and Research Challenges," \textit{2019 International Conference on Electrical and Computing Technologies and Applications (ICECTA)}, 2019, pp. 1-5.

53. E. A. Kadir, R. Shubair, S. K. Abdul Rahim, M. Himdi, M. R. Kamarudin and S. L. Rosa, "B5G and 6G: Next Generation Wireless Communications Technologies, Demand and Challenges," \textit{2021 International Congress of Advanced Technology and Engineering (ICOTEN)}, 2021, pp. 1-6.

54. S. Gong et al., "Toward Smart Wireless Communications via Intelligent Reflecting Surfaces: A Contemporary Survey," in \textit{IEEE Communications Surveys \& Tutorials}, vol. 22, no. 4, pp. 2283-2314, Fourthquarter 2020.

55. C. Pan et al., "Reconfigurable Intelligent Surfaces for 6G Systems: Principles, Applications, and Research Directions," in \textit{IEEE Communications Magazine}, vol. 59, no. 6, pp. 14-20, June 2021.

56. M. Di Renzo et al., "Reconfigurable Intelligent Surfaces vs. Relaying: Differences, Similarities, and Performance Comparison," in \textit{IEEE Open Journal of the Communications Society}, vol. 1, pp. 798-807, 2020.

57. B. Makki, K. Chitti, A. Behravan and M. -S. Alouini, "A Survey of NOMA: Current Status and Open Research Challenges," in \textit{IEEE Open Journal of the Communications Society}, vol. 1, pp. 179-189, 2020.

58. B. Zheng, Q. Wu and R. Zhang, "Intelligent Reflecting Surface-Assisted Multiple Access With User Pairing: NOMA or OMA?," in \textit{IEEE Communications Letters}, vol. 24, no. 4, pp. 753-757, April 2020.

59. M. Alsabah et al., "6G Wireless Communications Networks: A Comprehensive Survey," in \text{IEEE Access}, vol. 9, pp. 148191-148243, 2021.

60. S. Kumar et al., “A survey on IRS NOMA integrated communication networks,” in \textit{Telecommunication Systems}, vol. 80, pp. 277–302, April 2022.

61. H. Wang, C. Liu, Z. Shi, Y. Fu and R. Song, "On Power Minimization for IRS-Aided Downlink NOMA Systems," in \textit{IEEE Wireless Communications Letters}, vol. 9, no. 11, pp. 1808-1811, Nov. 2020.

62. Y. Cheng, K. H. Li, Y. Liu and K. C. Teh, "Outage Performance of Downlink IRS-Assisted NOMA Systems," \textit{GLOBECOM 2020 - 2020 IEEE Global Communications Conference}, 2020, pp. 1-6.

63. M. Fu, Y. Zhou and Y. Shi, "Intelligent Reflecting Surface for Downlink Non-Orthogonal Multiple Access Networks," \textit{2019 IEEE Globecom Workshops (GC Wkshps)}, 2019, pp. 1-6.

64. J. Zhu, Y. Huang, J. Wang, K. Navaie and Z. Ding, "Power Efficient IRS-Assisted NOMA," in \textit{IEEE Transactions on Communications}, vol. 69, no. 2, pp. 900-913, Feb. 2021.

65. A. S. de Sena et al., "IRS-Assisted Massive MIMO-NOMA Networks with Polarization Diversity," \textit{2021 IEEE International Conference on Communications Workshops (ICC Workshops)}, 2021, pp. 1-6.

66. J. Zuo, Y. Liu, Z. Qin and N. Al-Dhahir, "Resource Allocation in Intelligent Reflecting Surface Assisted NOMA Systems," in \textit{IEEE Transactions on Communications}, vol. 68, no. 11, pp. 7170-7183, Nov. 2020.

67. P. Liu, Y. Li, W. Cheng, X. Gao and X. Huang, "Intelligent Reflecting Surface Aided NOMA for Millimeter-Wave Massive MIMO With Lens Antenna Array," in \textit{IEEE Transactions on Vehicular Technology}, vol. 70, no. 5, pp. 4419-4434, May 2021.

68. Z. Ding, R. Schober and H. V. Poor, "On the Impact of Phase Shifting Designs on IRS-NOMA," in \textit{IEEE Wireless Communications Letters}, vol. 9, no. 10, pp. 1596-1600, Oct. 2020.

69. W. Tang et al., "Wireless Communications With Reconfigurable Intelligent Surface: Path Loss Modeling and Experimental Measurement," in \textit{IEEE Transactions on Wireless Communications}, vol. 20, no. 1, pp. 421-439, Jan. 2021.

70. M. TEKBAŞ, A. TOKTAŞ and G. ÇAKIR, "Design of a Dual Polarized mmWave Horn Antenna Using Decoupled Microstrip Line Feeder," \textit{2020 International Conference on Electrical Engineering (ICEE)}, 2020, pp. 1-4.
}

\end{document}